\documentclass[usenatbib]{mn2e}
\usepackage[dvips]{graphicx,color}
\usepackage{amssymb}
\title[An Overview of Long-term Variability in X-ray Binaries]
  {Characterizing X-ray Binary Long-term Variability}
\author[M.M. Kotze \& P.A. Charles]
  {M.M.~Kotze $^{1,2}$,
  P.A.~Charles $^{1,2,3}$\\
  $^1$ South African Astronomical Observatory,
       P.O. Box 9, Observatory 7935, South Africa (SA)\\
  $^2$ Astrophysics, Cosmology and Gravity Centre (ACGC), Astronomy Department, University of Cape Town, Rondebosch 7701, SA\\
  $^3$ School of Physics \& Astronomy, University of Southampton, Southampton SO17 1BJ, UK}
\date{Released 2011}

\pagerange{\pageref{firstpage}--\pageref{lastpage}} \pubyear{2011}

\def\LaTeX{L\kern-.36em\raise.3ex\hbox{a}\kern-.15em
    T\kern-.1667em\lower.7ex\hbox{E}\kern-.125emX}

\begin{document}
\label{firstpage}

\maketitle

\begin{abstract}
\noindent 
Long term (``superorbital'') periods or modulations have been detected in a wide variety of both low and high-mass X-ray binaries at X-ray and optical wavelengths. A variety of mechanisms have been proposed to account for the variability properties, such as precessing and/or warped accretion discs, amongst others. The All Sky Monitor on board the Rossi X-ray Timing Explorer provides the most extensive ($\sim$15 years) and sensitive X-ray archive for studying such behaviour. It is also clear that such variations can be intermittent and/or a function of X-ray spectral state. Consequently, we use a time-dependent Dynamic Power Spectrum method to examine how these modulations vary with time in 25 X-ray binaries for which superorbital periodicities have been previously reported. Our aim is to characterize these periodicities in a completely systematic way. Some (such as Her X-1 and LMC X-4) are remarkably stable, but others show a range of properties, from even longer variability time-scales to quite chaotic behaviour.

\end{abstract}
\bigskip
\section{Introduction} 
The Galactic population of X-ray binaries (XRBs) are X-ray luminous, interacting binaries in which a donor star ($M_{2}$) transfers material to a neutron star (NS) or black hole (BH) compact object ($M_{X}$). Low mass X-ray binaries (LMXBs) contain donor stars $<1 M_{\odot}$ and typically have orbital periods ($P_{orb}$) ranging from hours to days (Liu, van Paradijs \& van den Heuvel 2007), while high mass X-ray binaries (HMXBs) have donor stars $>10 M_{\odot}$ and $P_{orb}\sim$ several days to tens of days (Liu, van Paradijs \& van den Heuvel 2006). 

From the very first decade of satellite X-ray astronomy (the 1970s), long-term monitoring revealed modulations that were quasi-periodic on time-scales substantially longer than their well-established orbital periods. These included results from missions such as Ariel V's All Sky Monitor (e.g. Kaluzienski et al. 1976) and Vela 5B (e.g. Priedhorsky \& Terrell 1983, 1984). Her X-1, with its 35 d on/off cycle (which is remarkably stable, at $\sim 20\times P_{orb}$) exhibits the prototypical ``superorbital'' period (Petterson 1977).

\subsection{Long-term variations} 

Such superorbital variations ($P_{sup}$) are seen to occur on time-scales of tens to hundreds of days, and are thought to be related to the properties of the accretion disc (and occasionally linked to the donor). Many of these have been determined for a number of sources, using archival data from later X-ray satellite missions such as CGRO's BATSE (e.g. Robinson et al. 1997), RXTE's All Sky Monitor (ASM; e.g. Wen et al. 2006) and Swift's Burst Alert Telescope (e.g. Farrell, Barret \& Skinner 2009). Most recently, superorbital behaviour and the various mechanisms proposed to account for it, have been summarised in Charles et al. (2008, 2010). 

Radiation-induced warping (Ogilvie \& Dubus 2001, hereafter OD01) and/or precession (Whitehurst \& King 1991) of the accretion disc may lead to periodic/quasi-periodic superorbital modulation of the X-ray flux, e.g. Cyg X-2, Her X-1, LMC X-4 \& SMC X-1 (Clarkson et al. 2003). However, superorbital variations may also occur as the result of other mechanisms. Here we briefly introduce the physical processes that have been proposed.

\subsubsection{Disc Precession} 
Precession of accretion discs may occur if $q<$0.25-0.33, due to resonances excited in the discs resulting from tidal interaction with the donor (Whitehurst \& King 1991). This occurs in the SU UMa sub-class of Cataclysmic Variables (CVs; Warner 1995) and the soft X-ray Transients (SXTs; O'Donoghue \& Charles 1996), containing white dwarfs (WD) and BH compact objects respectively. It is very likely also to occur in ultra-compact X-ray binaries (UCXBs; e.g. X1916-053) where the donor itself is degenerate (Nelemans 2004).

\begin{figure}
  \centering
  \includegraphics[angle=0,width=0.45\textwidth]{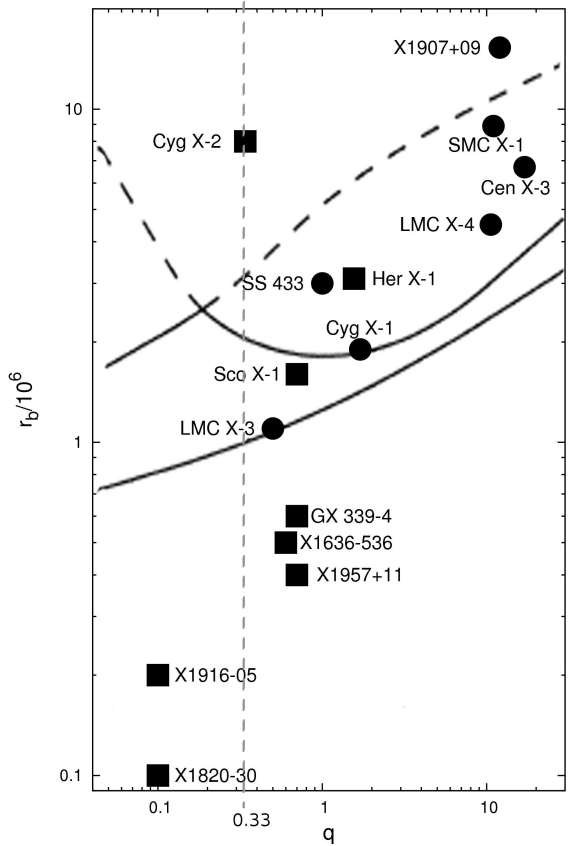}
  \caption {OD01 stability analysis for radiation-driven warping in X-ray binaries, using $\alpha=0.3$ and $\epsilon=0.1$ (adapted from Fig.7 in OD01, including only sources with known $P_{sup}$), where $r_{b}$ has units $[GM_{X}/c^{2}]$ and q is dimensionless. The upper two lines (solid \& dashed) correspond to the first two bending modes for $r_{add}=r_{c}$ (mode 0 starting higher than mode 1). Sustained stable warps are only possible close to the solid portions of those curves, while warps become increasingly variable/unstable as the dashed curves are approached and crossed. The instability zone lies between these solid and dashed curves. The bottom solid line corresponds to mode 1 for $r_{add}=r_{o}$, below which discs are unlikely to experience radiation-driven warping. The region between the bottom and top solid lines are considered an intermediate instability zone where a disc may display warping cycles as it alternates between being warped and being flat. Squares indicate LMXBs and circles HMXBs. The grey vertical dashed line indicates the $q\sim$0.33 boundary (Whitehurst \& King 1991), to the left of which sources are susceptible to disc precession resulting from tidal interactions with the donor.} 
  \label{OD2001}
\end{figure}

\subsubsection{Radiation-induced Warping/Tilting} 

This is the original superorbital mechanism that was proposed to account for the 35d cycle in Her X-1 by Petterson (1977). Subsequently developed by Wijers and Pringle (1999), the most detailed analysis to date is that of OD01. They showed how accretion discs can develop radiation-induced warps that are functions of their binary separation ($r_{b}$) and mass ratio ($q=M_{2}/M_{X}$). The stability of such warps is defined in this analysis by the range of values for the mass input radius ($r_{add}$), with boundary values set by the circularization radius ($r_{c}$) and the outer disc radius ($r_{o}$). (See Fig.\ref{OD2001} for further details.) 

OD01 ignored tidal torque and assumed a viscosity parameter ($\alpha=0.3$) and accretion efficiency ($\epsilon=0.1$) that were considered reasonable for persistent X-ray binaries in general. It is expected that chaotic (e.g. Cyg X-2) or unstable (e.g. SMC X-1) warping may result in quasi-periodic superorbital variations, while precession of stable/sustained warps (e.g. Her X-1 \& LMC X-4) should result in steady, almost strictly periodic superorbital modulations.

\subsubsection{Magnetic Warping} 

Pfeiffer \& Lai (2004) carried out numerical simulations of viscous accretion discs driven by the magnetic torque in the inner disc, resulting from the interaction between the induced electric currents in the disc and the accreting object's magnetic field ($\overrightarrow{B}$). They show that an accretion disc surrounding a magnetic star (T Tauri star, WD or NS) can develop warps in the inner disc that steadily precess. The retrograde precession as observed in Her X-1, is successfully reproduced by their simulations. They suggest that changes in $\dot{M}$ can cause discs to switch between warped and flat.

\subsubsection{Wind-driven Tilting} 

Schandl \& Meyer (1994) proposed an alternative mechanism for the tilted disc observed for Her X-1, involving the coronal wind torque in combination with viscous and tidal torques, wherein the coronal wind exerts a repulsive force on the disc (but see also the analysis and discussion concerning the coronal model in Scott, Leahy \& Wilson (2000)). Quillen (2001) proposed that a supersonic wind passing over the accretion disc may cause surface instabilities. Moreover, Montgomery \& Martin (2010) suggest it may also cause lift due to the differing speeds above and below the disc, causing a pitch of the disc around its line of nodes, thereby tilting it. 

\subsubsection{X-ray State Changes} 

Long-term modulation of the mass transfer rate ($\dot{M}$) through their discs has been associated with dramatic ``state changes'' in the X-ray spectra of many sources (of which Cyg X-1 and GX339-4 have provided frequent demonstrations). This is particularly true for the SXTs (McClintock \& Remillard 2006 and references therein) and luminous galactic bulge sources (van der Klis 2006 and references therein). In such cases, the mass transfer rate is intermediate between the stable high and low spectral states, and so occasional transitions will occur. Super-orbital periods detected in such systems may therefore indicate the time-scales associated with such transitions (see e.g. King et al. 1997).

Quasi-persistent XRBs (e.g. KS 1731-260 \& X1636-536) can display prolonged high (on) and low (off) intervals, characterized by markedly different flux levels during those times (Charles et al. 2008). During the decline from the high to the low state, X1636-536 displayed an anti-correlation between the hard and soft X-rays as well as superorbital variations consistent with modulations of $\dot{M}$ comparable with the viscous time-scale in the outer disc (Shih et al. 2005, 2011). 

\subsubsection{Precessing Relativistic Jets} 

The archetypal galactic micro-quasar is the well-known SS433 (Hjellming \& Johnston 1981; Margon 1984), which displays the remarkable ``moving lines'' in its optical spectrum with radial velocities that are truly relativistic. These are now well-established as arising from precessing relativistic jets, a process which imparts substantial stability to the 162 d superorbital period. The jets and their corkscrew motion have been directly (spatially) observed at radio wavelengths (Blundell, Bowler \& Schmidtobreick 2007), although the detailed properties of the 13 d orbital period XRB at the heart of SS433 remain poorly determined (e.g. Barnes et al. 2006). The driving mechanism for the jet precession in SS433 is still a matter of controversy, but irradiation driven outflows are likely to play a significant role, as discussed in e.g. Begelman et al (2006).

\subsubsection{Third Body} 

The most luminous LMXB in a galactic globular cluster is X1820-303 (at the core of NGC 6624). With close to the shortest orbital period ($\sim$12 mins), it should not be susceptible to radiation-induced disc warping as suggested by OD01. But its very low $q$ (the donor is a very low mass degenerate helium dwarf), does make it a prime candidate for tidal disc precession effects. Nevertheless, the extremely stable superorbital period which has been observed in X1820-303 is considered a likely consequence of the effect of a third body on the $\dot{M}$ in the system (Chou \& Grindlay, 2001). Such triple systems involving an XRB have long been mooted, especially within the extremely dense stellar regions of globular cluster cores, but this system remains the strongest candidate.

\subsubsection{Be X-ray Binaries} 

When a NS in a highly eccentric orbit passes through, or interacts with, the varying equatorial disc surrounding a rapidly rotating Be-star, it naturally gives rise to a modulation at the orbital frequency (see e.g. Leahy \& Kostka 2008 and references therein, and  Hayasaki \& Okazaki 2006). However, there are longer term variations in which the equatorial disc expands and contracts which will give rise to a modulation of the accreted fraction on this superorbital time-scale (Charles et al. 2008). These are the Be X-ray binaries (BeX), which display a wide variety of variations at X-ray \& optical wavelengths, extending to time-scales much longer than being considered here (Rajoelimanana, Charles \& Udalski 2011).

\subsection{Time-dependent Period Analysis} 
Periods are typically identified by applying a periodic analysis to the entire available dataset to produce a periodogram. While this approach allows one to distinguish between periodic (sharp peaks) and quasi-periodic (broadened or multiple peaks) behaviour, it is essentially limited to the identification (not characterization) of periodic signals (Wen et al. 2006).

However, from our brief introduction to the variety of physical mechanisms that can induce superorbital variations, it will be clear that some of these can be intermittent, or vary in time-scale so much as to be rendered undetectable in a summed periodogram. Consequently, such behaviour should be analysed in a way that displays the ``instantaneous'' modulations present in the data, and then follow this as a function of time.

A time-dependent periodic analysis of two HMXBs and two LMXBs in the form of Dynamic Power Spectra (DPS; e.g. Clarkson et al. 2003), clearly illustrated the need for a more comprehensive approach in the study of superorbital periods. Therein the $P_{sup}$ in SMC X-1 was seen to evolve dramatically and systematically, while those in Her X-1 and LMC X-4 remained essentially steady in period, although the modulation in Her X-1 would occasionally disappear completely e.g. Still et al. (2004). In contrast, the $P_{sup}$ observed in Cyg X-2 appeared to be highly variable and unstable.

The DPS technique is well established and the X-ray burst community uses a variant of it routinely (e.g. Fig.19 in Watts et al. (2005)). However, since it requires significantly more computations than the typical period analysis, its application was limited in the past. Recent advances in desktop technology allow this technique to be employed more readily than before.

Another approach to unravel the time-dependence of these long-term periods was recently developed by Hu et al. 2011, and this is the Hilbert-Huang Transform (HHT). Hu et al applied the HHT to the same SMC X-1 dataset as used here, and find that the HHT is consistent with results obtained from the DPS. They suggest that it may be able to show more detail in the frequency and time domains.

Boyd \& Smale (2004) applied time-frequency decomposition techniques to the long-term RXTE ASM lightcurves of three XRBs, separating them into random and periodic components. They suggest that all the $P_{sup}$ modulations in Cyg X-2 are integer multiples of its $P_{orb}$, although they concede that competing warping modes (suggested in OD01) could be considered as one of several mechanisms which may possibly be responsible for its long-term variations.

\subsection{Target Details} 

With an RXTE observational baseline that has doubled, we revisit the 4 sources of Clarkson et al. (2003). In addition, we apply the same analysis to the other 14 sources showing $P_{sup}$ behaviour and summarised in Charles et al. (2008). A further 7 sources, not contained therein or published since, were also included in our target list.

Here we present a time-resolved periodic analysis of all 25 sources with previously published or claimed $P_{sup}<$1 yr, many of which have been inferred to be associated with warped and/or precessing accretion discs. Consequently, we endeavour to provide a comprehensive overview of the long-term behaviour of the majority of superorbital modulations observed in X-ray binaries, through this systematic DPS approach. In this way we hope that these results can provide a fresh context within which superorbital periods in X-ray binaries can be investigated.

The 9 HMXBs and 16 LMXBs considered here are listed in Tables \ref{HMXBs} \& \ref{LMXBs} respectively, together with their previously published $P_{sup}$, $P_{orb}$ and $q=\frac{M_2}{M_X}$ (values of which are taken from OD01 unless otherwise stated). Their positions (if known) are shown on an adapted OD01 plot (Fig.\ref{OD2001}) of instability zones with respect to $r_{b}$ (in units of $[GM_{X}/c^{2}]$) and $q$.


\begin{table}
	\caption{HMXBs with known superorbital periods}
	\label{HMXBs}
	\smallskip
	\begin{tabular}{lccc}
	\hline
	\bf{Source} &  \bf{$P_{sup}$ [d]}  &  \bf{$P_{orb}$ [d]} &  \bf{$q$}\\
	\hline
	Cen X-3  & 140  $^{[O]}$ & 2.09 $^{[L]}$ &  17.0 \\  
	Cyg X-1  & 142 $^{[O]}$, 326  $^{[R]}$ & 5.6 $^{[L]}$ &  1.7 \\
	   &  150 \& 290  $^{[La]}$ &  &  \\
	LMC X-3  & 99 $^{[C2]}$, 100-500 $^{[W]}$ &  1.70 $^{[C1]}$ &  0.5 \\
	LMC X-4  & 30  $^{[W]}$  &  1.41 $^{[W]}$ &  10.6 \\
	SMC X-1  & 50-70  $^{[W]}$ & 3.89 $^{[S]}$ &  11.0 \\
	SS433 &  162  $^{[W]}$  & 13.10 $^{[C]}$ &  [1.0] \\
	X0114+650  & 31  $^{[W]}$  & 11.6 $^{[L]}$ &  \\
	X1907+097  & 42  $^{[P]}$  & 8.38 $^{[L]}$ &  12.0 \\
	XTE J1716-389  & 99  $^{[W]}$  & &  \\
	\hline
	\end{tabular}	
	\smallskip \\
	\footnotesize{$^{[C1]}$ Cowley et al. (1983),} 
	\footnotesize{$^{[C2]}$ Cowley et al. (1991),} \\
	\footnotesize{$^{[C]}$ Crampton, Cowley \& Hutchings (1980),} \\
	\footnotesize{$^{[O]}$ OD01 is the only reference found for it,} \\
	\footnotesize{$^{[La]}$ Lachowicz et al. (2006),} \\
	\footnotesize{$^{[L]}$ Liu, van Paradijs \& van den Heuvel (2006) and ref. therein,}\\
	\footnotesize{$^{[P]}$ Priedhorsky \& Terrell (1984),}\\
        \footnotesize{$^{[R]}$ Rico (2008),} \\
        \footnotesize{$^{[S]}$ Schreier et al. (1972),} \\	
        \footnotesize{$^{[W]}$ Wen et al. (2006) and ref. therein} \\
\end{table}

\begin{table}
	\caption{LMXBs with known superorbital periods}
	\label{LMXBs}
	\smallskip
	\begin{tabular}{lccc}
	\hline
	\bf{Source}  & \bf{$P_{sup}$ [d]} &  \bf{$P_{orb}$ [d]}  &  \bf{$q$}\\
	\hline  
	Cyg X-2  & 50-80 $^{[C]}$, 60-90 $^{[W]}$ & 9.844 $^{[L]}$ &  0.34\\
	EXO 0748-676  & 181  $^{[K]}$  & 0.158  $^{[L]}$ &  \\
	GRS 1747-312  & 147  $^{[W]}$  & 0.515  $^{[L]}$ &  \\
	GX 339-4  & 190-250  $^{[C]}$  & 1.755 $^{[L]}$ &  [0.7] \\
	GX 354-0  & 63 or 72  $^{[L]}$  & &  \\
	Her X-1  & 33-37  $^{[Le]}$  & 1.700 $^{[L]}$ &  1.56 \\
	IGR J17098-3628  & 163  $^{[K]}$  & &  \\
	KS 1731-260  & 38  $^{[C]}$  & &  \\
	LMC X-2  & 8  $^{[C]}$  & 0.34 $^{[L]}$ &  \\
	MS 1603.6+2600  & 5  $^{[C]}$  & 0.077 $^{[L]}$ &  \\
	Sco X-1  &  2.6 $^{*}$ & 0.788 $^{[L]}$ &  [0.7] \\
	X1636-536  & 46  $^{[C]}$  & 0.158 $^{[L]}$ &  [0.6] \\
	X1730-333  & 217  $^{[W]}$  & &  \\
	X1820-303  & 172  $^{[W]}$  & 0.008 $^{[L]}$ &  [0.1] \\
	X1916-053  & 5  $^{[C]}$ \& 199  $^{[P]}$  & 0.035 $^{[L]}$ &  [0.1] \\
	X1957+115  & 117  $^{[C]}$  & 0.390 $^{[L]}$ &  [0.7] \\
	\hline
	\end{tabular}	
	\smallskip \\
        \footnotesize{$^{[C]}$ Charles et al. (2008) and ref. therein,}\\
	\footnotesize{$^{[K]}$ Kotze, Charles \& Crause (2009),}\\
        \footnotesize{$^{[L]}$ Liu, van Paradijs \& van den Heuvel (2007) and ref. therein,}\\
	\footnotesize{$^{[Le]}$ Leahy \& Igna (2010),}\\
	\footnotesize{$^{[P]}$ Priedhorsky \& Terrell (1984),}\\
        \footnotesize{$^{[W]}$ Wen et al. (2006) and ref. therein,}\\
        \footnotesize{$^{*}$ Quoted as $\sim$62 d (OD01), not $\sim$62 hrs (Kudryavtsev et al. 1989)}\\
\end{table}

\section{Observations and Data Analysis}

\subsection{RXTE/ASM}

The Massachusetts Institute of Technology (MIT) has operated an All Sky Monitor (ASM) on board the Rossi X-ray Timing Explorer (RXTE) since early 1996. The ASM observes the X-ray sky by using 3 rotating Scanning Shadow Cameras (SSC) to scan $\sim$80\% of the sky during each $\sim$90 min orbit. Data are reduced and compiled weekly by the ASM team and made publicly available \footnote{http://xte.mit.edu/ASMlc.html} as dwell-by-dwell or one-day-averages in four energy bands, which includes a Sum-band (1.5-12 keV). One-day-average data are the dwell-by-dwell data binned into 1-day bins. Archival ASM datasets contain the lightcurves of all X-ray sources in the RXTE catalogue. For full details see Levine et al. (1996) and http://xte.mit.edu/.

\subsection{DPS Variability Analysis}

The ASM one-day-average datasets used in this paper span from 20 February 1996 to 12 February 2011 (MJD 55608) and therefore provide 15 year light-curves of unparalleled quality and sensitivity. While period analysis of the entire dataset can provide the maximum sensitivity to low-level modulations (if they are steady), it has serious limitations when dealing with unstable or evolving periodic signals. Signals that are intermittent may be completely damped out, and evolving signals may cause multiple peaks over a much larger frequency range than is expected for so-called quasi-periodic oscillations (see e.g. van der Klis 2006). These effects get worse with longer observational baselines. Since superorbital periods relating to accretion disc behaviour are expected to be quasi-periodic, or could be unstable and may drift systematically, our interest rather lies in characterising the behaviour of those periods over time.

\subsubsection{Windowing}

The Dynamic Power Spectrum (DPS) method employed by Clarkson et al. (2003) has the advantage of illustrating very clearly whether periodic signals are sustained, and whether they are stable or drifting in period (although the method is also sensitive to phase drifts; Clarkson et al. 2003). The method requires the datasets to be split into windows that are of sufficient length to allow detection of the maximum period considered. A Lomb-Scargle (L-S) periodogram (Lomb 1976 \& Scargle 1982, 1989) was produced for every such dataset and the results of all the periodograms for a source were plotted together in a density map, using the L-S power for each frequency plotted at every window's mid-point along the time-axis. The frequency domain covered periods with a minimum of 2 d to a maximum comparable to the window size. 

Larger window sizes cause the smearing out of variability as it averages out small variations, but it has the advantage of enhancing sustained periodic signals. However, variable or intermittent periodic signals may be completely washed out if windows are too large. Window sizes were initially chosen to be 400 d for all sources, since it provides coverage of $\sim$5-10 periodic cycles in the range $\sim$40-80 d and it has the advantage of being directly comparable to the earlier results already published for SMC X-1, LMC X-4, Her X-1 \& Cyg X-2 (Clarkson et al. 2003). It also allowed us to identify longer potential periods, which could then be investigated in more detail by choosing more appropriate window sizes to sample them properly. For periods $>$80 d, windows of $\sim$5 times the periods were used for the results presented here, and for periods $<$10 d, we used 100 d windows.

A sliding window approach was also employed, whereby consecutive datasets overlap such that they move in 50 d steps in the time domain. This approach provides adequate temporal resolution, whilst smoothing out the noise from otherwise independent windows that will occur. 

\subsubsection{Dynamic Window Function}

To ensure that the features contained in the DPS are not artificially induced by the sampling or windowing employed, the spectral window function was determined for each dataset and a density map was constructed with that information. The resulting Dynamic Window Function (DWF) was compared to the DPS for each source. No significant features contained in the DPS were repeated in the DWF of any source considered here. 

\subsubsection{Dwell-by-dwell data}

The DPS analysis was also repeated for 1-day averages that were constructed from the dwell-by-dwell data, employing aggressive filtering of the data to eliminate possibly dubious data points, as suggested on the RXTE ASM website. Those results are consistent with what we present here. However, the data loss due to the more conservative approach, does lead to diminished temporal coverage. We find that the features in the DPS are in general clearer using the ASM 1-day averages, which allows better coverage, and are therefore the results presented here.

\subsubsection{Noise Levels}

White noise (frequency independent) levels were determined for each complete 15 year one-day-average dataset by generating $10^{4}$ random datasets by Monte Carlo simulation, assuming a white noise distribution and using the time values, mean and standard deviation of the dataset. Lomb-Scargle periodograms for each random dataset yield periods associated with the maximum power in each, and the subsequent distribution of powers amount to a cumulative probability distribution function, providing the false alarm probability (result of white noise) associated with a particular power. The white noise level for a 99.99\% confidence level, is the power associated with a false alarm probability of 0.01\%. The noise level for the dataset as a whole can be used as an estimate for its windows and is not expected to differ significantly between sources.

Red noise (frequency dependent) levels for the RXTE ASM data, determined by Farrell, Barret \& Skinner (2009) for 5 of these sources, indicated that the red noise levels do not always exceed the white noise at the low frequencies. For Sco X-1 and X1916-053 the white noise levels were $\sim$double those of the red noise, while Cyg X-2, X1636-536 \& X1820-303 had red noise levels that rose to $>$3 times the white noise levels at low frequencies (0.001-0.01). Apart from differing significantly for each source, these modelled red noise levels are extremely dependent on the time domain covered by the dataset, as well as the weighting scheme applied to it, if any. 

However, the aim of this paper is to characterize the behaviour of previously reported periodic signals over time, not claim the discovery of new periodic behaviour. The latter would normally require rigorous testing against noise, while such requirements should be irrelevant in case of the former. While noise levels may serve as an accepted statistical measure for the significance of prominent periodic signals, the true measure of their significance lies in their persistence over time, which a time-dependent periodic analysis provides automatically.

\subsection{Results}

To allow easier interpretation of features in the power density maps (the 2-D representation of 3-D data) and how they translate to features contained in the lightcurves, the DPS (from the Sum-band one-day-average data) are plotted, together with the lightcurve of each source. The light-curves shown were re-binned into 10-day bins and plotted above the DPS, over the same temporal range (horizontal/x-axis). 

Furthermore, the L-S periodogram for the entire 15 year dataset was plotted to the left of the DPS for each source and over the same frequency range (vertical/y-axis). This allows immediate comparison of features in the DPS to the overall periodogram. Frequency labels were replaced by the equivalent period labels, to facilitate easier comparison with previously published $P_{sup}$. 

Known $P_{sup}$ contained in Tables \ref{HMXBs} \& \ref{LMXBs} are mentioned in the caption of each source and are also clearly indicated on each plot with ticks along the period/frequency-axis (red in the online version of this paper). The density scale-bar included in each plot indicates the power (line-of-sight/z-axis) associated with the periodogram, with stronger detections being darker. 

DPS and L-S plots were produced for frequencies appropriate to their reported superorbital periods. Including a wider range of frequencies around those periods allows a more complete picture of the temporal properties of each source. In an effort to allow comparability between sources, the size of the frequency range plotted was kept relatively constant at $\sim$0.02-0.03 for all the plots, except that sources with $P_{sup}$ $<$10 d were plotted over a frequency range of 0.1. We only include the plot for the $P_{sup}\sim$199 d in X1916-053 (Fig.\ref{x1916}), since we find no evidence for the $P_{sup}\sim$5 d.

The estimated white noise levels for all sources are, not surprisingly, quite similar, being in the range from 15.6-16.8; they are plotted as vertical lines (green in the online version of this paper) on the L-S plots of the sources as a guide. Prominent features in the L-S periodogram which are well above these estimated white noise levels, can normally be considered significant, the situation for the majority of the sources considered here. Only in Figs. \ref{x1907}, \ref{lmcx2} \& \ref{ms1603} were no significant features detected in the L-S or DPS at any time.

Our aim here is to present the results in a way that provides the reader with sufficient information to reach their own conclusion regarding the variability properties of a source at a single glance. All relevant additional information, not contained in the plots themselves, was therefore included in the caption of each source. The figures for the sources are presented in the same order as their listing in Tables \ref{HMXBs} \& \ref{LMXBs}, with the HMXBs presented before the LMXBs.

While the results presented here certainly include most of the X-ray binaries with published superorbital periods, we do not claim to have included all sources for which superorbital periods have been published, nor do we claim to have included all X-ray binaries that have ever displayed superorbital behaviour. However, those included here do have superorbital periods that have been proposed to be associated with warped and/or precessing accretion discs at some stage. Their inclusion is intended as an opportunity to study such features using the DPS approach, thereby indicating whether such modulations were steady, evolving, persistent or intermittent.



\begin{figure}
  \centering
	\includegraphics[angle=0,width=0.47\textwidth]{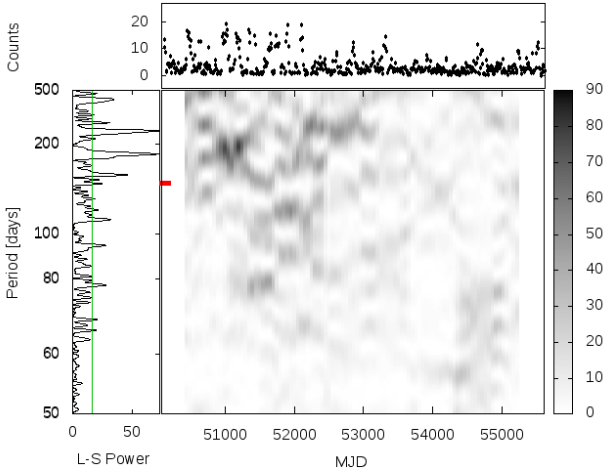}
  \caption {Cen X-3 : HMXB with $P_{sup}\sim$140 d (in OD01 only), indicated by marker on y-axis. Top panel: RXTE ASM lightcurve, Left panel: L-S over entire $\sim$15 yr dataset, Main panel: DPS with time vs period (x-y), Right: scale bar indicating DPS power (z).} 
  \label{cenx3}
\end{figure}

\begin{figure}
  \centering
	\includegraphics[angle=0,width=0.47\textwidth]{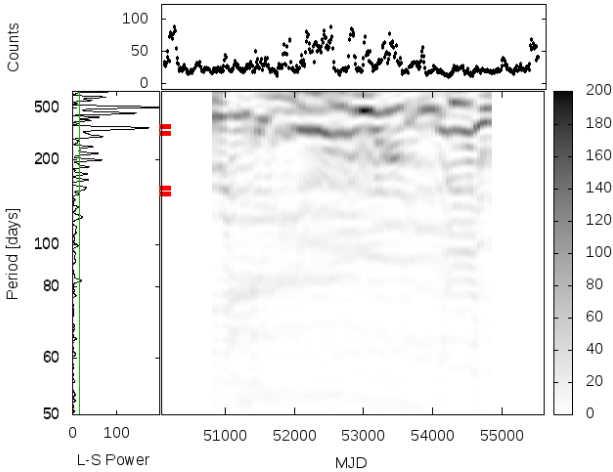}
  \caption {Cyg X-1 : HMXB with $P_{sup}\sim$142 d (in OD01 only), $\sim$150 \& 290 d (Lachowicz et al. 2006) and $\sim$326 d (Rico 2008).} 
  \label{cygx1}
\end{figure}

\begin{figure}
  \centering
	\includegraphics[angle=0,width=0.47\textwidth]{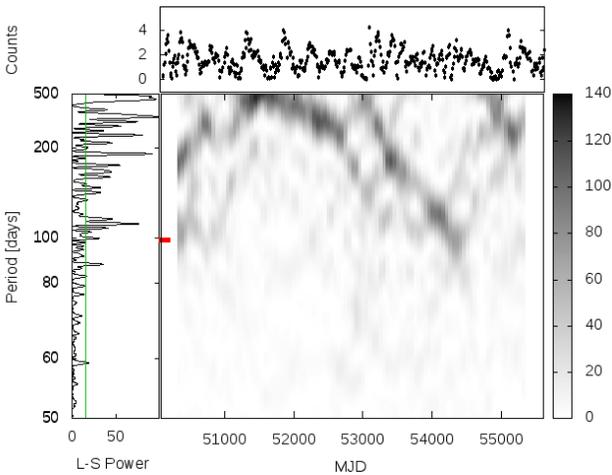}
  \caption {LMC X-3 : HMXB with $P_{sup}\sim$99 d (Cowley et al. 1991).} 
  \label{lmcx3}
\end{figure}

\begin{figure}
  \centering
	\includegraphics[angle=0,width=0.47\textwidth]{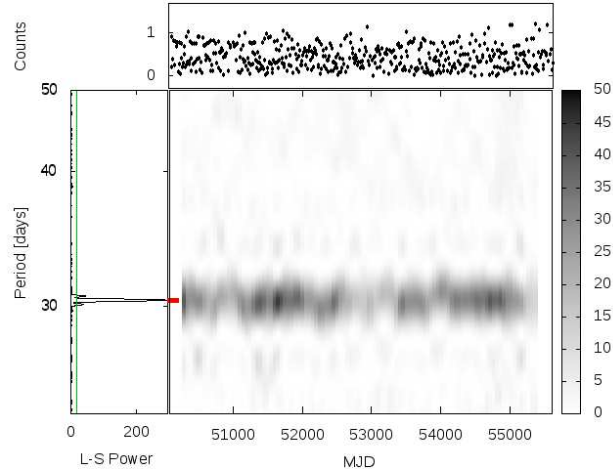}
  \caption {LMC X-4 : HMXB with $P_{sup}\sim$30 d (Wen et al. 2006).} 
  \label{lmcx4}
\end{figure}

\begin{figure}
  \centering
	\includegraphics[angle=0,width=0.47\textwidth]{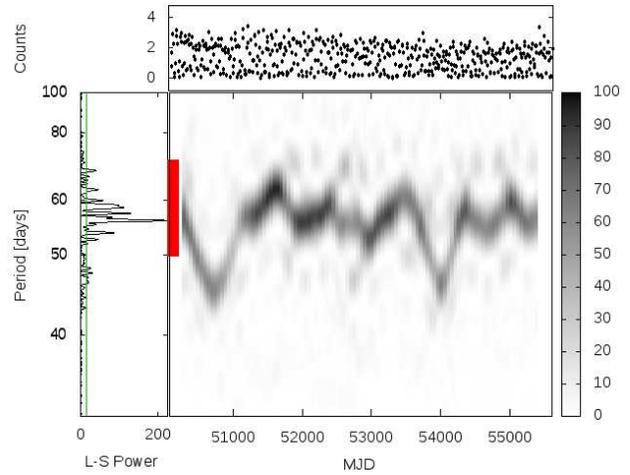}
  \caption {SMC X-1 : HMXB with $P_{sup}\sim$50-70 d (Wen et al. 2006).} 
  \label{smcx1}
\end{figure}

\begin{figure}
  \centering
	\includegraphics[angle=0,width=0.47\textwidth]{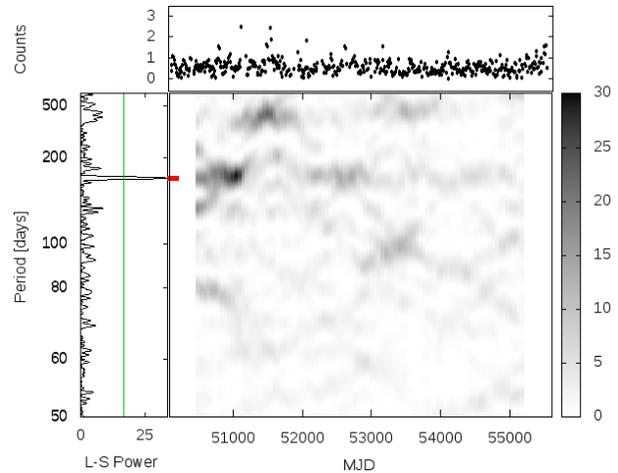}
  \caption {SS433 : HMXB with $P_{sup}\sim$162 d (Wen et al. 2006).} 
  \label{ss433}
\end{figure}

\clearpage

\begin{figure}
  \centering
	\includegraphics[angle=0,width=0.47\textwidth]{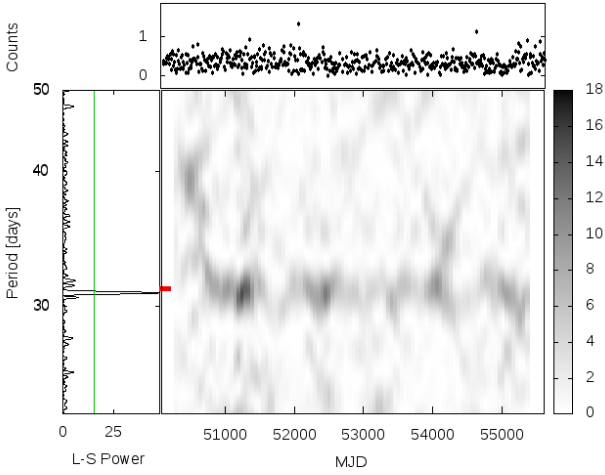}
  \caption {X0114+650 : HMXB with $P_{sup}\sim$31 d (Wen et al. 2006).} 
  \label{x0114}
\end{figure}

\begin{figure}
  \centering
	\includegraphics[angle=0,width=0.47\textwidth]{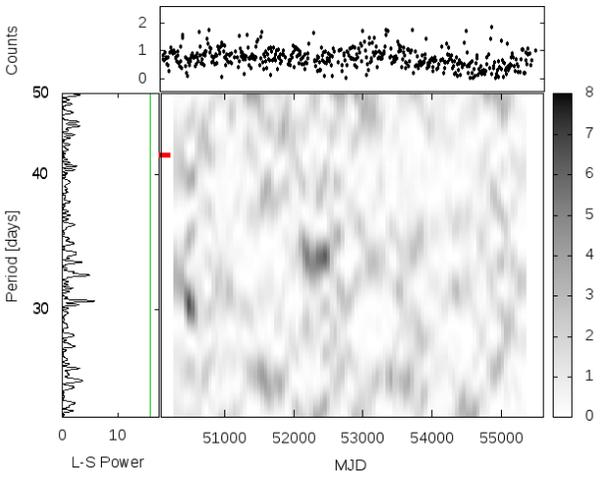}
  \caption {X1907+097 : HMXB with $P_{sup}\sim$42 d (Priedhorsky \& Terrell 1984). All features are below white noise.} 
  \label{x1907}
\end{figure}

\begin{figure}
  \centering
	\includegraphics[angle=0,width=0.47\textwidth]{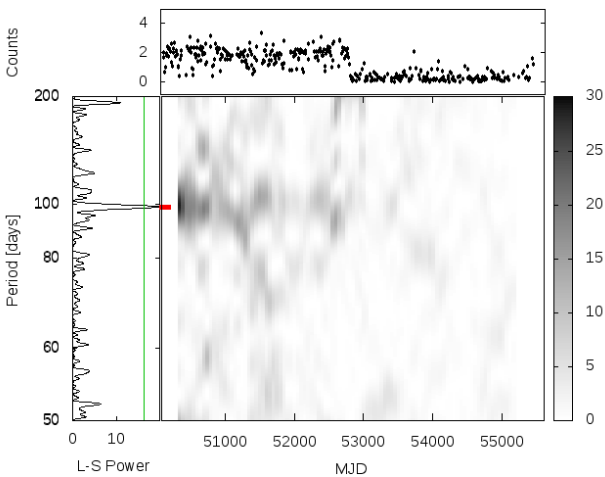}
  \caption {XTE J1716-389 : HMXB with $P_{sup}\sim$99 d (Wen et al. 2006).} 
  \label{xtej1716}
\end{figure}


\begin{figure}
  \centering
	\includegraphics[angle=0,width=0.47\textwidth]{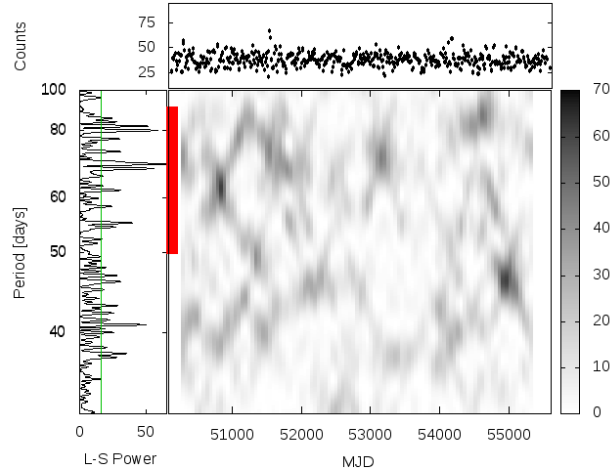}
  \caption {Cyg X-2 : LMXB with $P_{sup}\sim$50-80 d (Charles et al. 2008) or 60-90 d (Wen et al. 2006).} 
  \label{cygx2}
\end{figure}

\begin{figure}
  \centering
	\includegraphics[angle=0,width=0.47\textwidth]{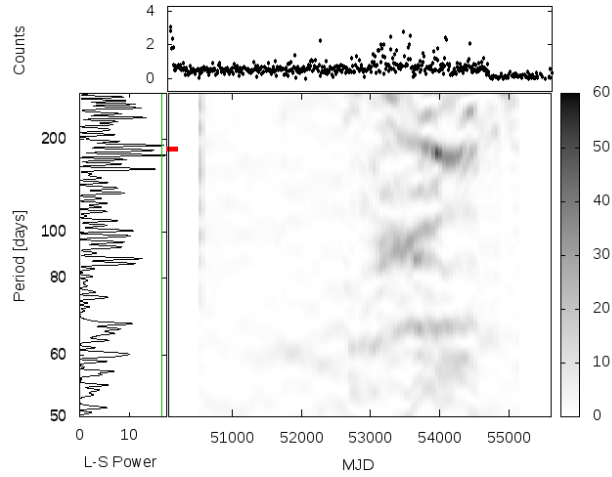}
  \caption {EXO 0748-676 : LMXB with $P_{sup}\sim$181 d during higher state (Kotze, Charles \& Crause 2009).} 
  \label{exo0748}
\end{figure}

\begin{figure}
  \centering
	\includegraphics[angle=0,width=0.47\textwidth]{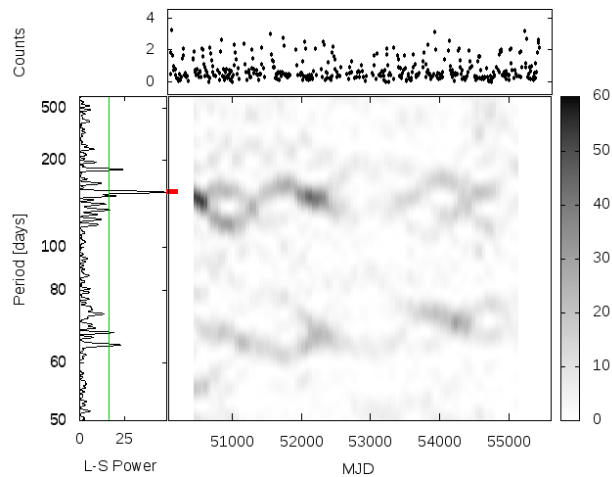}
  \caption {GRS 1747-312 : LMXB with $P_{sup}\sim$147 d (Wen et al. 2006).} 
  \label{grs1747}
\end{figure}

\clearpage

\begin{figure}
  \centering
	\includegraphics[angle=0,width=0.47\textwidth]{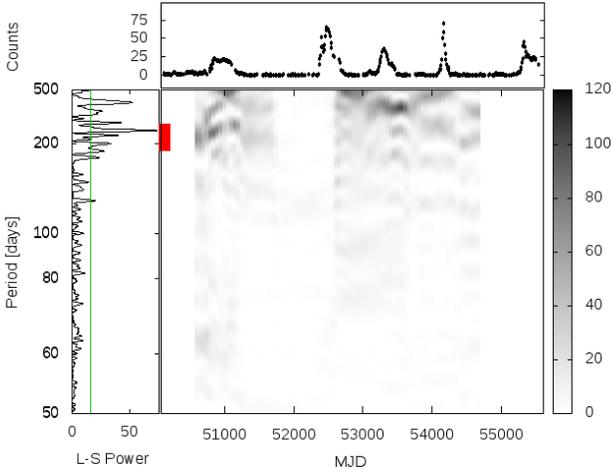}
  \caption {GX 339-4 : LMXB with $P_{sup}\sim$190-250 d (Charles et al. 2008).} 
  \label{gx339}
\end{figure}

\begin{figure}
  \centering
	\includegraphics[angle=0,width=0.47\textwidth]{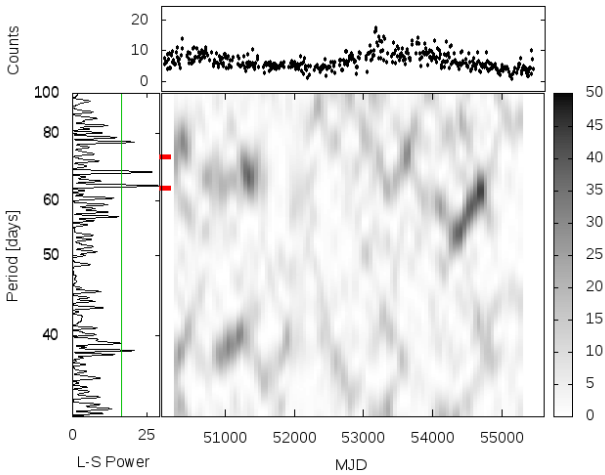}
  \caption {GX 354-0 : LMXB with $P_{sup}\sim$63 or 72 d (Liu, van Paradijs \& van den Heuvel 2007).} 
  \label{gx354}
\end{figure}

\begin{figure}
  \centering
	\includegraphics[angle=0,width=0.47\textwidth]{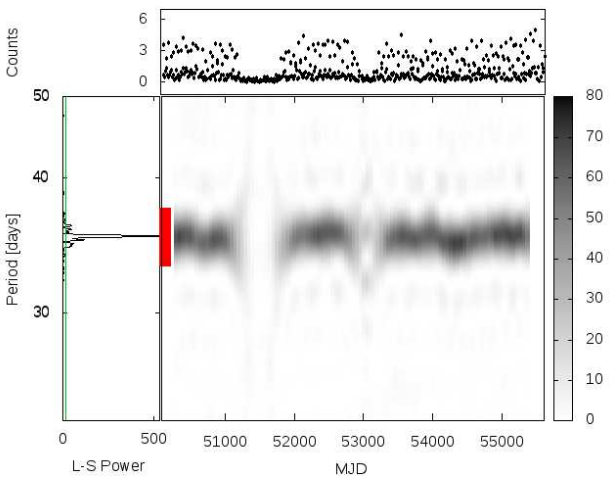}
  \caption {Her X-1 : LMXB with $P_{sup}\sim$33-37 d, with average $\sim$35 d (Leahy \& Igna 2010).} 
  \label{herx1}
\end{figure}

\begin{figure}
  \centering
	\includegraphics[angle=0,width=0.47\textwidth]{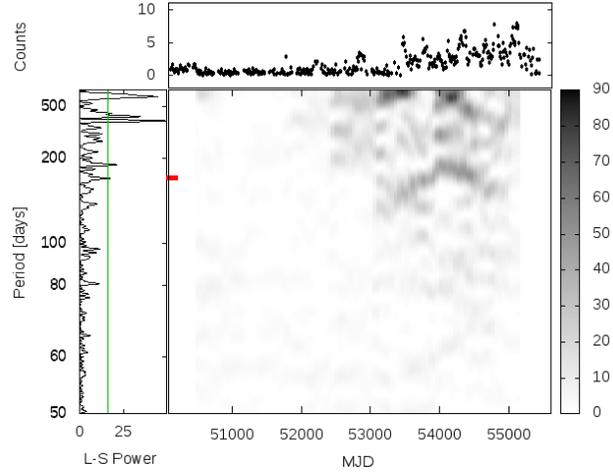}
  \caption {IGR J17098-3628 : LMXB with $P_{sup}\sim$163 d during part of the high state (Kotze, Charles \& Crause 2009).}
  \label{igr17098}
\end{figure}

\begin{figure}
  \centering
	\includegraphics[angle=0,width=0.47\textwidth]{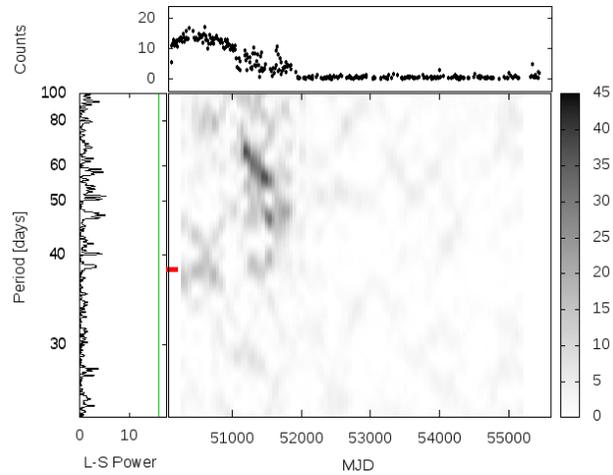}
  \caption {KS 1731-260 : LMXB with $P_{sup}\sim$38 d (Charles et al. 2008).} 
  \label{ks1731}
\end{figure}

\begin{figure}
  \centering
	\includegraphics[angle=0,width=0.47\textwidth]{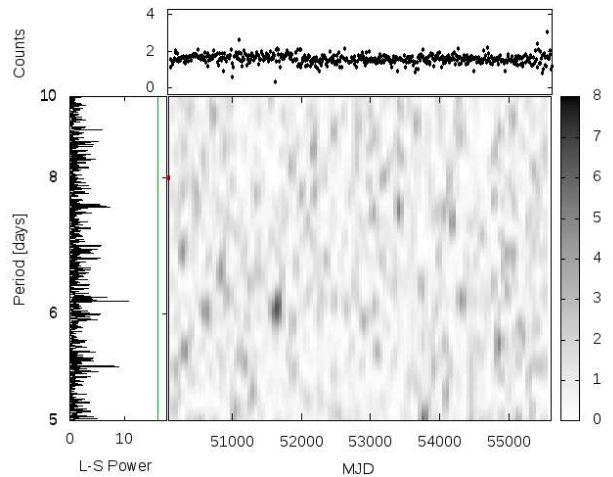}
  \caption {LMC X-2 : LMXB with $P_{sup}\sim$8 d (Charles et al. 2008). All features are below white noise.} 
  \label{lmcx2}
\end{figure}

\clearpage

\begin{figure}
  \centering
	\includegraphics[angle=0,width=0.47\textwidth]{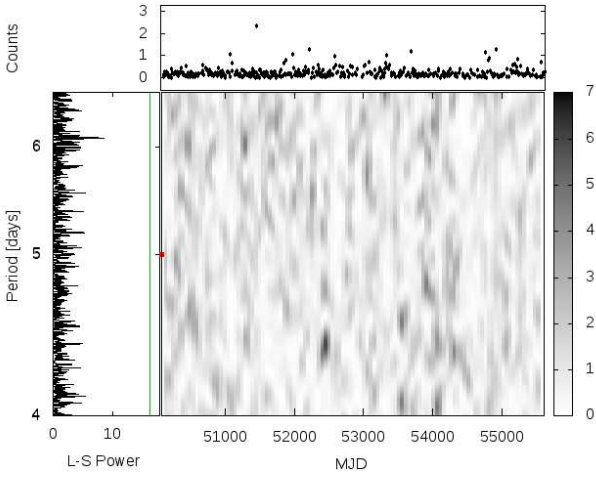}
  \caption {MS 1603.6+2600 : LMXB with $P_{sup}\sim$5 d (Charles et al. 2008). All features are below white noise.} 
  \label{ms1603}
\end{figure}

\begin{figure}
  \centering
	\includegraphics[angle=0,width=0.47\textwidth]{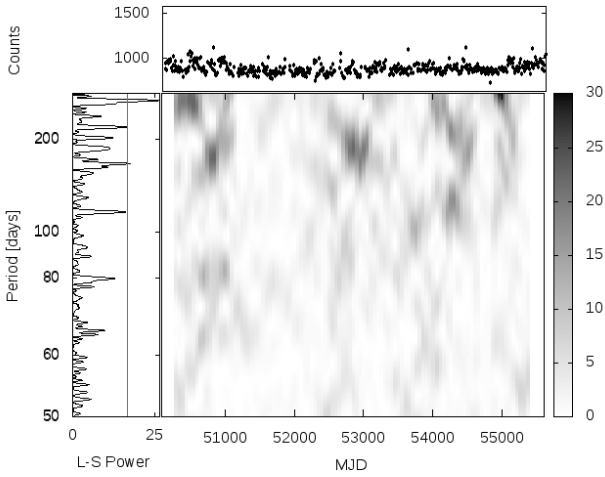}
  \caption {Sco X-1 : LMXB with $P_{sup}\sim$2.6 d (Kudryavtsev et al. 1989).} 
  \label{scox1}
\end{figure}

\begin{figure}
  \centering
	\includegraphics[angle=0,width=0.47\textwidth]{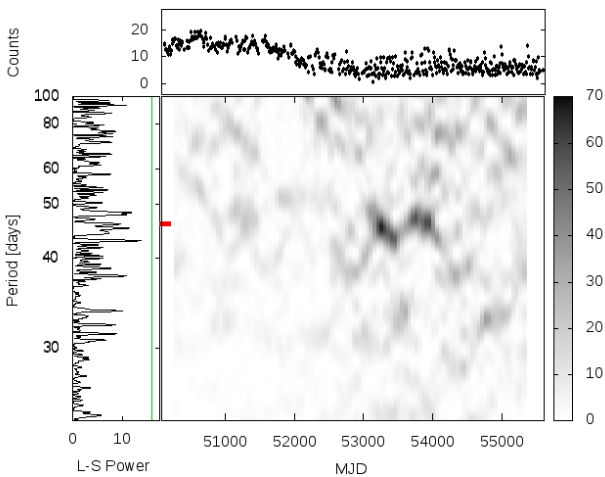}
  \caption {X1636-536 : LMXB with $P_{sup}\sim$46 d after transition to lower state at MJD$\sim$52000 (Charles et al. 2008).} 
  \label{x1636}
\end{figure}

\begin{figure}
  \centering
	\includegraphics[angle=0,width=0.47\textwidth]{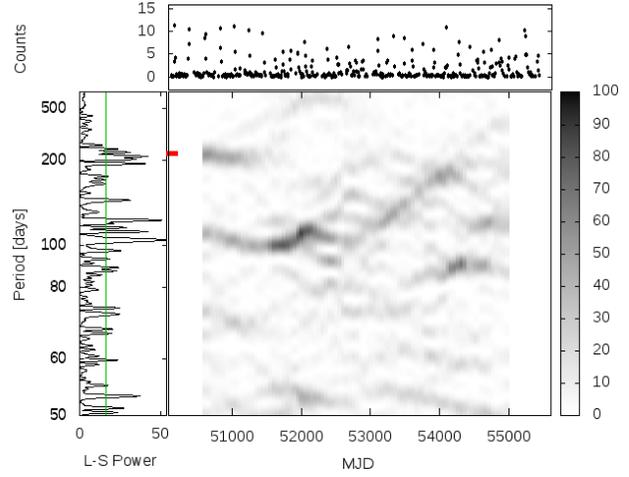}
  \caption {X1730-333 : LMXB with $P_{sup}\sim$217 d (Charles et al. 2008).} 
  \label{x1730}
\end{figure}

\begin{figure}
  \centering
	\includegraphics[angle=0,width=0.47\textwidth]{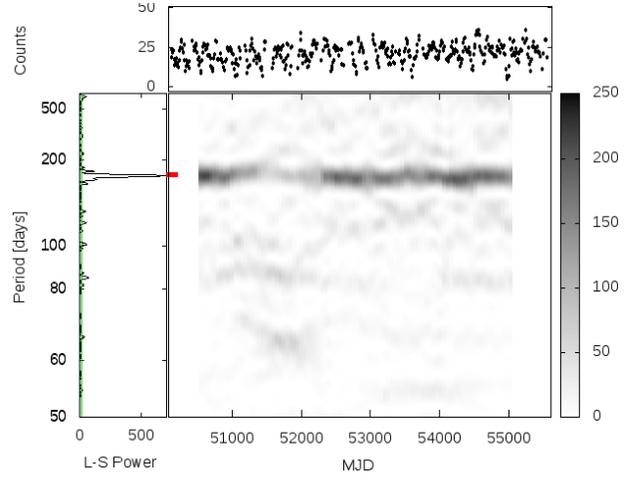}
  \caption {X1820-303 : LMXB with $P_{sup}\sim$172 d (Wen et al. 2006).} 
  \label{x1820}
\end{figure}

\begin{figure}
  \centering
	\includegraphics[angle=0,width=0.47\textwidth]{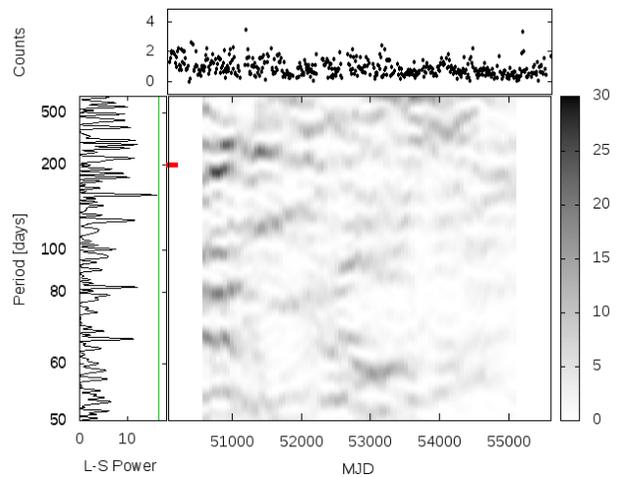}
  \caption {X1916-053 : LMXB with $P_{sup}\sim$199 d (Wen et al. 2006).} 
  \label{x1916}
\end{figure}

\clearpage

\begin{figure}
  \centering
	\includegraphics[angle=0,width=0.47\textwidth]{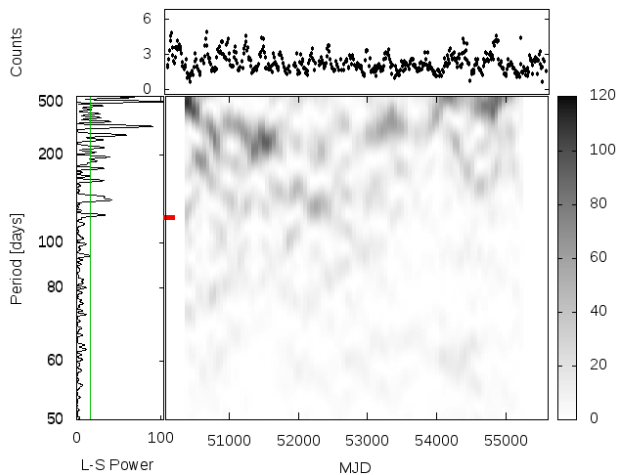}
  \caption {X1957+115 : LMXB with $P_{sup}\sim$117 d (Charles et al. 2008).} 
  \label{x1957}
\end{figure}

\section{Discussion}

It is only fitting that we begin our discussion with the up-to-date results for the 4 sources included in Clarkson et al. (2003). We will then discuss the remainder of those in Charles et al. (2008) plus 7 additional sources.  For each source we explicitly discuss its stability to radiation-driven warping under the OD01 framework. We remind the reader that we are {\it not} claiming that radiation-driven warping must explain superorbital periodicities for all sources.

Determining the behaviour of the previously reported superorbital periods required a time-dependent method, as suggested in OD01. They predicted that systems able to produce steady superorbital periods resulting from stable, steadily precessing, warped discs (as in Her X-1), should be rare, and variable/chaotic warping would result in a variety of quasi-periodic or aperiodic behaviour in the majority of sources. Our results here show that this prediction is largely borne out. While Fig.\ref{OD2001} provides us with the OD01 stability predictions for individual sources, some were discussed in more detail in OD01 and where relevant we refer to those comments, since they pertain to deviations from the predictions implied in the figure.

\subsection{The Clarkson et al. (2003) sources}

Our DPS results for Her X-1, SMC X-1, LMC X-4 and Cyg X-2 are consistent with those presented in Clarkson et al. (2003) and those included in the Charles et al. (2008) review. However, the observational baseline used here has approximately doubled. These four sources therefore represent good examples with which the long-term behaviour of the superorbital periods of other sources can be compared, particularly in terms of OD01 disc instability.

\subsubsection{SMC X-1    (Fig.\ref{smcx1})}
Still by far the most intriguing of all our results, the DPS of this HMXB clearly shows the smooth and steady evolution of its $P_{sup}$ over the range $\sim$42-70 d. The behaviour early in the light-curve (the feature surrounding MJD$\sim$50500), also included in Clarkson et al. (2003) and Trowbridge, Nowak \& Wilms (2007), now appears to be repeated approximately 10 years later (around MJD$\sim$54000). There also appear to be a decrease in maximum and increase in minimum 10-day binned flux coincident with the temporal location of the minimum superorbital periods in those two features in the DPS curve. The behaviour after these two largest features does not repeat, but appears to be similar in terms of the decreased period range covered, and contains features that seem to recur on a time-scale of $\sim$1000 d. 

SMC X-1 lies in the instability zone on Fig.\ref{OD2001}. However, it is situated closer to the dashed curve and therefore a radiation-induced warp in its accretion disc is expected to be variable, though not yet unstable. The steady evolution of its superorbital period may subsequently be the result of the precession of a steadily warped disc in which the warp itself varies, due to the competition of warping modes (Clarkson et al. 2003). Full investigation and further interpretation of these results warrants a far more detailed discussion of the source itself that goes beyond the scope of this paper.

\subsubsection{LMC X-4    (Fig.\ref{lmcx4})}
Perhaps the steadiest of the stable superorbital periods in all HMXBs, a substantial modulation is strongly detected with a $P_{sup}$ of 30 d. The detection level appears to drop temporarily at MJD$\sim$51000, MJD$\sim$53000 and finally almost disappears completely around MJD$\sim$55500. Since the source is clearly detected throughout our observational baseline, the disappearance of the periodic signal toward the end is not due to poor coverage, but more likely an effect intrinsic to the source itself.

LMC X-4 lies in the instability zone on Fig.\ref{OD2001}, closer to the solid curve. The radiation-induced warps in its accretion disc are therefore expected to be stable and consequently able to produce a steady modulation.

\subsubsection{Her X-1    (Fig.\ref{herx1})}
This prototypical LMXB has continued to show a very strong detection of a relatively stable $P_{sup}\sim$35 d throughout the ASM coverage (and whenever it has been monitored over the last 40 years). The interruptions occurring around MJD$\sim$51250-51750 and MJD$\sim$53000, coincide with the anomalous low states (ALS), of which the first was included and discussed as such in Still et al. (2001). During these states, the maximum flux drops dramatically and becomes comparable to the minimum flux, which also appears to be decreasing slightly. Multi-wavelength data suggest that the ALS is not due to $\dot{M}$ being cut off, but is rather caused by the shadowing effect of a more extreme tilt in the disc, which blocks the X-ray emission from the inner regions from the observer (Vrtilek et al. 2001). Leahy \& Dupuis (2010) suggest that a larger disc warp may account for the larger reduction in extreme UV flux (using Extreme Ultraviolet Explorer data) in comparison to that in X-rays (using simultaneous RXTE ASM data) during the ALS.

Her X-1 lies in the instability zone on Fig.\ref{OD2001}, approximately halfway between the solid and dashed curves. The superorbital period is therefore expected to be relatively stable, as a result of the precession of a radiatively warped accretion disc. However, it may show some evolution due to small-scale variation in the structure of the warp itself (see Leahy \& Igna 2010).

\subsubsection{Cyg X-2    (Fig.\ref{cygx2})}
Still showing chaotic and unstable behaviour, the range of peaks from $\sim$40-90 d in this LMXB are strongly detected from time to time throughout the RXTE lifetime. An initially weaker (but significant) detection in the range $\sim$40-50 d becomes a much stronger detection toward the end of the light-curve. 

Cyg X-2 lies well above the dashed curve on Fig.\ref{OD2001}, and is expected to experience chaotic warping of its accretion disc. As such, it is not expected to have a stable or steadily evolving superorbital period, but rather intermittent and variable periods produced by precessing unstable warps. The DPS is consistent with intermittent variations.

\subsection{Other HMXBs}

\subsubsection{Cen X-3    (Fig.\ref{cenx3})}
There is a detection of the $P_{sup}\sim$140 d around MJD$\sim$51500 for this HMXB, but it seems to form part of an evolving feature in the DPS, that had its strongest detection at $\sim$200 d at MJD$\sim$51000. A variety of shorter term signals (probably linked to the time-scale of the ``on'' states) are strongly detected up to MJD$\sim$53500, after which the maximum flux drops dramatically.

Cen X-3 is located in the OD01 instability zone in Fig.\ref{OD2001}, between LMC X-4 and SMC X-1. Consequently, Cen X-3 is expected to develop a warped precessing accretion disc in which the warp itself might be variable (Priedhorsky \& Terrell 1983; Iping \& Petterson 1990). The behaviour of the superorbital period is therefore also expected to be somewhere between stable (like LMC X-4) and steadily evolving (like SMC X-1). However, our DPS shows the evolution to be erratic, but not chaotic (like Cyg X-2).

\subsubsection{Cyg X-1    (Fig.\ref{cygx1})}
In this HMXB we detect the very long term $P_{sup}$ of $\sim$150 d and $\sim$290 d, which were detected in the Ariel 5 and Vela 5B datasets respectively (Lachowicz et al. 2006), and the subsequently detected $\sim$326 d period in the Swift BAT and RXTE ASM data (Rico 2008). In fact, apart from those we detect several other features, many of which appear to be stable and are not necessarily harmonics. 

Cyg X-1 is located on the OD01 curve associated with $r_{add}=r_{c}$ on Fig.\ref{OD2001}, where we expect stable warps to develop resulting in a steady superorbital period, like that in LMC X-4. Zdziarski, Pooley \& Skinner (2011) reported the doubling of the $P_{sup}$ and the constant shape of the intrinsic spectrum, concluding that those modulations are likely produced by a precessing accretion disc and not changes in $\dot{M}$. They also confirm the presence of a bulge on the accretion disc edge, as previously proposed.

\subsubsection{LMC X-3    (Fig.\ref{lmcx3})}
Features in the DPS of this HMXB are closely linked to details visible in the lightcurve, since the range of superorbital periods and the clearly visible lightcurve features are all $\sim$100-500 d. The periodic signals are therefore rather indicative of time-scales than periodicity (OD01). 

There is an initial evolution from shorter to longer term features, then a change of direction around MJD$\sim$51500 which continues past MJD$\sim$53000 and then collapses. However, a new long term cycle then starts, evolves more rapidly to a shorter term feature during MJD$\sim$53000-54500 and also appears to collapse, making way for another long term cycle starting at MJD$\sim$54700. We see no evidence for a steady $P_{sup}\sim$99 d as reported in the past (Cowley et al. 1991), but rather several steadily evolving long-term features covering the range $\sim$ 100-500 d, which is the range reported in Wen at al. (2006). 

LMC X-3 is located on the OD01 curve using $r_{add}=r_{o}$ in Fig.\ref{OD2001}, below which systems are not expected to develop warped precessing accretion discs. However, OD01 suggest that the long-term variations in this source are due to changes in the mass accretion rate from the donor. 

\subsubsection{SS433    (Fig.\ref{ss433})}
The strongest detection for this HMXB is $P_{sup}\sim$162 d, which is the well-known relativistic jet precession period (Margon et al. 1984). This disappears shortly after MJD$\sim$51000 and reappears at MJD$\sim$52000 at a weaker detection strength, remaining until MJD$\sim$55000. The weaker detections at the same frequency over the entire baseline and the singular strong detection in the overall L-S, suggest that a stable X-ray period $\sim$162 d is definitely present. The detection thereof may just be hampered from time to time by intrinsic fluctuations. A strong detection of a period $\sim$400 d at MJD$\sim$51000-52000, when the detection of the $\sim$162 d period effectively disappears, makes a weaker reappearance at MJD$\sim$53000-54000. During that time it is mirrored (on the opposite side of the $\sim$162 d period) by an equally weakly detected $\sim$100 d period, which appears to be unstable but persists to MJD$\sim$55000. Furthermore, upon its reappearance at MJD$\sim$51000-52000, the $\sim$400 d period appears to have evolved to an even longer period. Since the $\sim$400 d period is detected more strongly when the $\sim$162 d period is weakly detected, this might suggest that the presence of the longer period may have something to do with the weaker detections of the seemingly stable $\sim$162 d period.

SS433 is located in the instability zone on Fig.\ref{OD2001} next to Her X-1. A stable superorbital period is expected and is very clearly present in its optical and radio behaviour. The relativistic jets as inferred from their optical emission components, have been stable on the 162 d period for $>$ 30 yrs (Blundell \& Bowler 2005). We note that SS433 is in a very different evolutionary state at present compared to any other XRB considered here, with its high $q$ value leading to extreme super-Eddington accretion onto the BH (Begelman, King \& Pringle 2006).

\subsubsection{X0114+650    (Fig.\ref{x0114})}
In this HMXB, the $P_{sup}\sim$31 d is weakly detected from MJD$\sim$50500 in the DPS. It is strongly detected in the overall L-S periodogram and from MJD$\sim$50700-51400 in the DPS. Thereafter it remains relatively stable with alternating stronger and weaker detections of this signal. 

Although it is not as strongly detected as those of LMC X-4 or Her X-1, it appears relatively steady and persistent. The location of X0114+650 on Fig.\ref{OD2001} is not known, but as an HMXB with $P_{orb}\sim$ 11.6 d (Liu, van Paradijs \& van den Heuvel 2006) it is likely to be above the OD01 curve for $r_{add}=r_{o}$, and stable superorbital periods associated with a stable precessing warped accretion disc are therefore not unexpected. Moreover, since X0114+650 is a BeX (Koenigsberger et al. 1983), we could also expect superorbital variations due to a modulation in $\dot{M}$ as the equatorial disc of the Be-star expands and contracts (see Rajoelimanana, Charles \& Udalski 2011). 

\subsubsection{X1907+097    (Fig.\ref{x1907})}
We do not find evidence for the claimed (OD01) $P_{sup}\sim$42 d (or any other) X-ray superorbital period in this HMXB. X1907+097 is located above the dashed curve on Fig.\ref{OD2001} where we expect chaotic warping to occur, but OD01 suggested that it may lack superorbital variations as a result of its eccentric orbit (Bildsten et al. 1997).

\subsubsection{XTE J1716-389    (Fig.\ref{xtej1716})}
XTE J1716-389 is a quasi-persistent HMXB. An initial detection of the $P_{sup}\sim$99 d, becomes progressively weaker and is only visible in the DPS until MJD$\sim$53000, when there is a dramatic drop in the lightcurve to the low state.

The relatively regular, weaker detections of the $\sim$99 d period while the source is persistent, do seem to suggest that it exhibits a relatively steady superorbital period. The location of XTE J1716-389 on Fig.\ref{OD2001} is not known, but as an HMXB it is expected to be at least above the OD01 curve for $r_{add}=r_{o}$. As such, a stable superorbital period associated with a steadily precessing warp while it is persistent, is also not unexpected.

\subsection{Other LMXBs}

\subsubsection{EXO 0748-676    (Fig.\ref{exo0748})}
A $P_{sup}$ of 181 d was detected during the high state in this quasi-persistent LMXB and was clearly not stable, but shows steady evolution. The DPS shows several significant detected signals during the portions of the higher state where the previously published superorbital periods were noted (Kotze, Charles \& Crause 2009) and there is no evidence for any periodic signals in the low state. 

EXO 0748-676 has an unknown location on Fig.\ref{OD2001}. However, it has $P_{orb}\sim$3.8 hours (Liu, van Paradijs \& van den Heuvel 2007) and $q\sim$0.11-0.28 (Mu{\~n}oz-Darias et al. 2009), implying $M_{2}<$0.42. As a LMXB with $P_{orb}<$1 day, we would not expect it to produce a steadily precessing warped accretion disc, but with a $q<$0.25-0.33, tidal disc precession is a serious possibility.

\subsubsection{GRS 1747-312    (Fig.\ref{grs1747})}
The $P_{sup}\sim$147 d of this LMXB is strongly detected in the L-S periodogram and DPS at MJD$\sim$50500 \& MJD$\sim$52000. However, this periodic signal shows clear variation that appears to continue through an interval with lower amplitude variations (significantly weaker detections around MJD$\sim$52500-53500), to re-emerge thereafter. 

There is also a significant detection at $\sim$70 d, which appears to be related to (but not simply the harmonic of) the periodic behaviour around $\sim$147 d. The two periodic signals at $\sim$70 and $\sim$147 d both appear to evolve overall toward slightly longer periods. 

The location of GRS 1747-312 on Fig.\ref{OD2001} is not known, but as a LMXB with $P_{orb}\sim$12.36 hours (Liu, van Paradijs \& van den Heuvel 2007) it is likely situated below the OD01 curve for $r_{add}=r_{o}$, below which sources are not expected to develop warped precessing discs or display the steady $P_{sup}$ associated therewith.

\subsubsection{GX 339-4    (Fig.\ref{gx339})}
The $P_{sup}\sim$190-250 d in this transient-like BH-LMXB are detected in the lightcurve after removing the outbursts (the plot included herein) as well as when the outbursts are included. GX 339-4 is located below the curve for $r_{add}=r_{o}$ on Fig.\ref{OD2001} and is not expected to develop a warped precessing disc and variations are likely due to modulations in $\dot{M}$ (OD01). However, the $q\sim$0.7 used in OD01 was a rough estimate and it has since been determined to be $q<$0.08 (Hynes et al. 2003), making tidal disc precession likely. Furthermore, Hynes et al. (2003) find $P_{orb}\sim$1.755 d and $M_{X}\sim$5$M_{\odot}$. Recalculation using the OD01 method gives $\frac{r_{b}}{10^{6}}\sim$1 in comparison with $\sim$0.6 in OD01, moving GX 339-4 into the intermediate instability zone.

\subsubsection{GX 354-0    (Fig.\ref{gx354})}
The DPS shows a $P_{sup}\sim$63 \& $\sim$72 d are both initially detected, and rapidly replaced by a single period between those values. The strongest detections are towards the end of the dataset where a $\sim$50 d period evolves toward $\sim$70 d. The source also clearly displays very long-term variability on a $\sim$decade long time-scale (\"Ozdemir 2010, Kotze \& Charles 2010), which are not believed to be associated with the accretion disc properties. 

We find no evidence for a stable superorbital period, nor for prolonged steady evolution. The behaviour rather resembles the chaotic warping observed in Cyg X-2. Its location on Fig.\ref{OD2001} is unknown.

\subsubsection{IGR J17098-3628    (Fig.\ref{igr17098})}
A $P_{sup}$ of 163 d was detected during the higher state in this quasi-persistent LMXB. It is clearly not stable and shows steady evolution with several significant detected signals in the DPS during the portions of the higher state where the previously published superorbital periods were noted (Kotze, Charles \& Crause 2009). There is no evidence for periodic signals in the low state. 

IGR J17098-3628 has an unknown location on Fig.\ref{OD2001}, but in Kotze, Charles \& Crause (2009) we argued that, based on the apparent similarities with EXO 0748-676, IGR J17098-3628 likely also has $P_{orb}<$1 d. As a LMXB with $P_{orb}<$1 day, IGR J17098-3628 would not be expected to produce steadily precessing warped accretion discs (OD01).

\subsubsection{KS 1731-260    (Fig.\ref{ks1731})}
The $P_{sup}\sim$38 d of this quasi-persistent LMXB is detected during the high state (MJD$<$51000) and during the transition from the high state to the low state (MJD$\sim$51000-52000). However, the strongest detection is a period that evolves from $\sim$70 d to $\sim$45 d during that transition. So there appears to be two sections to the high state which could be considered separately, namely MJD$<$51000 and MJD$\sim$51000-52000. The $\sim$38 d superorbital period appears in both of those, suggesting some measure of stability. 

Although KS 1731-260 does not appear to be stable like Her X-1 when it is persistent, its behaviour is not quite comparable to the chaotic warping behaviour observed in Cyg X-2. KS 1731-260's location on Fig.\ref{OD2001} is unknown, but as an LMXB it is likely located below the OD01 curve for $r_{add}=r_{o}$ and therefore unlikely to produce a warped accretion disc.

\subsubsection{LMC X-2 \& MS 1603.6+2600    (Fig.\ref{lmcx2} \& \ref{ms1603})}
The DPS results for these LMXBs show no evidence for any significant long-term X-ray periodicities, particularly excluding the previously suggested $P_{sup}\sim$8 d and $\sim$5 d for LMC X-2 \& MS 1603.6+2600 respectively (Charles et al. 2008). All points in their L-S periodograms are below the white noise level and their DPS plots show no periodic features that are persistent in any way. We note that periods in the same range are very clearly detected for several other sources using similar data.

The behaviour for LMC X-2 and MS 1603.6+2600 is not even reminiscent of Cyg X-2, where there are significant, but chaotic signals that are persistent and strongly detected. The results for these two sources appear to represent only noise. 

Locations for both these sources are unknown on Fig.\ref{OD2001}. MS 1603.6+2600 has $P_{orb}\sim$1.85 hours and LMC X-2 has $P_{orb}\sim$8.16 hours (Liu, van Paradijs \& van den Heuvel 2007) and as LMXBs with $P_{orb}<$1 d, they are not expected to produce warped accretion discs (OD01). However, with $q\sim$0.14 for LMC X-2 (Cornelisse et al. 2007), tidal disc precession is likely. MS 1603.6+2600 a.k.a. UW CrB, has a short $P_{orb}$ and likely shares properties with the other short period systems, such as having a $q<$0.25, making tidal disc precession likely.

\subsubsection{Sco X-1    (Fig.\ref{scox1})}
The previously reported $P_{sup}$ was incorrectly quoted as $\sim$62 d in OD01, instead of $\sim$62 hrs (Kudryavtsev et al. 1989). Neither of these are significantly detected here. However, longer term periods $\sim$200 d are strongly detected and appear to recur after essentially disappearing from MJD$\sim$51000-52000.

Sco X-1 is located in the OD01 intermediate instability zone on Fig.\ref{OD2001}, where discs are expected to alternate between being warped and flat. However, the $q\sim$0.7 used in OD01 was only an estimate, and it has since been determined to be $q\sim$0.3 (Steeghs \& Casares 2002), making tidal disc precession also likely. Recalculation using the OD01 method gives $\frac{r_{b}}{10^{6}}\sim$1.5, meaning Sco X-1 remains in the intermediate instability zone. 

\subsubsection{X1636-536    (Fig.\ref{x1636})}
The $P_{sup}\sim$46 d superorbital period of this quasi-persistent LMXB shows similar behaviour to that of KS 1731-260 in that it is strongly detected at MJD$\sim$53000-54000 when the source has made a transition to a lower state (Shih et al. 2005). It does not appear to be stable during either state, but appears persistent and steadily evolving during the first half of the low state. The second part of the low state contains detections of shorter periods.

The location of X1636-536 on Fig.\ref{OD2001} makes formation of a warped precessing disc unlikely since it is below the $r_{add}=r_{o}$ curve. Warps are also expected to form during the persistent states of quasi-persistent X-ray binaries, but here the periodic signals are stronger during the low state. The quasi-periodic superorbital period detected in X1636-536 is believed to be a result of X-ray state changes, characterized by the anti-correlation between hard and soft X-rays (Shih et al. 2005). The optical and soft X-rays have been found to be correlated, as the latter is reprocessed in the accretion disc (Shih et al. 2011).

\subsubsection{X1730-333    (Fig.\ref{x1730})}
This prototypical LMXB (it is the Rapid Burster in the globular cluster Liller 1 (Apparao \& Chitre 1979)) shows an initially strong detection of the $P_{sup}\sim$217 d, which appears to remain steady up to MJD$\sim$51500. However, during that time there are a number of shorter term periods, that do not appear to simply be harmonics. A period of mostly $\sim$100 d is strongly detected and steadily drifting until MJD$\sim$52500, when a marked evolution from shorter to longer periods begins. 

The superorbital periods in X1730-333 show evolution but are not steadily evolving (in the SMC X-1 sense), neither are they unstable (in the Cyg X-2 sense). It rather appears to be a variety of steadily evolving periods, which initially appear to have some connection with each other. They may be indicative of the time-scale on which bursts recur, rather than periodicities. The location of X1730-333 on Fig.\ref{OD2001} is unknown.

\subsubsection{X1820-303    (Fig.\ref{x1820})}
The $P_{sup}\sim$172 d in this LMXB is strongly detected and appears stable. Its detection is weaker during MJD$\sim$51200-52200 (but still significant) and simultaneously there are detections of shorter term periods during that time. The $\sim$86 d period is the first harmonic thereof, but the $\sim$65 d period is not a harmonic.

The location of X1820-303 on Fig.\ref{OD2001}, makes it highly unlikely that any superorbital period in the system would be related to a precessing warped accretion disc. However, since it has been interpreted as a triple system (Chou \& Grindlay, 2001), the period would rather be connected with the $\dot{M}$ variations resulting from the effects of the third body and is therefore expected to be (extremely) stable. The minor variations visible in the DPS for the $P_{sup}\sim$172 d are within conservative error estimates for the periods determined in each window, and are therefore not significant.

\subsubsection{X1916-053    (Fig.\ref{x1916})}
There is an initial strong detection of the $P_{sup}\sim$199 d before MJD$\sim$51000, but we found no evidence for any modulation near 5 d. Several periodic signals are detected initially, but none that remains stable or show steady evolution.

As a double degenerate LMXB, the location of X1916-053 on Fig.\ref{OD2001} makes it highly unlikely that any superorbital period in the system would be related to a precessing warped accretion disc. However, with $q$ well below 0.25 (OD01) it would be very likely highly susceptible to disc precession due to tidal interactions with the donor (see e.g. Homer et al 2001).

\subsubsection{X1957+115    (Fig.\ref{x1957})}
The results for this LMXB show detection of a $P_{sup}\sim$117 d initially and at MJD$\sim$52250, which appears to form part of an unsteadily evolving feature in the DPS. There are longer term features that are more significantly detected, but they are also unstable. 

According to the location of X1957+115 on Fig.\ref{OD2001}, it is not expected to develop a warped precessing disc. OD01 suggest periodic signals may rather be a product of the time-scale of features in the lightcurve, due to changes in $\dot{M}$. 


\begin{table}
    \begin{center}
	\caption{Characterization of $P_{sup}$ behaviour in HMXBs}
	\label{hmxbs}
	\smallskip
	\begin{tabular}{lccc}
	\hline
	\bf{Source} &  \bf{Behaviour}  &  \bf{Mechanism} &  \bf{Types}$^{[L]}$ \\
	\hline
	Cen X-3  & varies & mode 0-1$^{*}$ & P,E,C \\  
	Cyg X-1  & steady & mode 0$^{*}$ & U,R \\
	LMC X-3  & varies  &  $\Delta \dot{M}$/$r_{add}=r_{o}$ $^{*}$ \\
	LMC X-4  & steady  &  mode 0$^{*}$ \\
	SMC X-1  & varies &  mode 0-1$^{*}$ \\
	SS433 &  steady  & jet/mode 0$^{*}$ \\
	X0114+650  & steady  & BeX & P,C? \\
	X1907+097  & --  &  mode 1+$^{*}$ & P,T,C \\
	XTE J1716-389  & steady  & unknown \\ 
	\hline
	\end{tabular}	
	\smallskip \\
	\footnotesize{$^{*}$ OD01}\\
	\footnotesize{$^{[L]}$ Liu, van Paradijs \& van den Heuvel (2006) and ref. therein,}\\
	\footnotesize{P: X-ray pulsar,}
	\footnotesize{T: transient X-ray source,}\\
	\footnotesize{E: eclipsing system,}
	\footnotesize{R: radio emitting HMXBs,}\\
	\footnotesize{C: cyclotron resonance scattering feature at X-ray spectrum,}\\
	\footnotesize{U: ultra-soft X-ray spectrum} \\
    \end{center}
\end{table}

\begin{table}
    \begin{center}
	\caption{Characterization of $P_{sup}$ behaviour in LMXBs}
	\label{lmxbs}
	\smallskip
	\begin{tabular}{lccc}
	\hline
	\bf{Source} &  \bf{Behaviour}  &  \bf{Mechanism} &  \bf{Types}$^{[L]}$ \\
	\hline
	Cyg X-2  & chaotic & mode 1+$^{*}$ & B,Z,R \\
	EXO 0748-676  & varies  & $\Delta \dot{M}$ & T,B,D,E  \\
	GRS 1747-312  & varies  & unknown & T,G,B,D,E \\
	GX 339-4  & varies  & $\Delta \dot{M}$ & T,U,M,R \\
	GX 354-0  & chaotic  & unknown & B,A,R \\
	Her X-1  & steady & mode 0$^{*}$ & P,D,E \\
	IGR J17098-3628  & varies  & $\Delta \dot{M}$ & T,R? \\
	KS 1731-260  & varies  & $\Delta \dot{M}$ & T,B \\
	LMC X-2  &  --  & --  & Z \\
	MS 1603.6+2600  & --  & -- & B \\
	Sco X-1  &  chaotic & intermediate$^{*}$  & Z,M,R \\
	X1636-536  & varies  & $\Delta \dot{M}$ & B,A \\
	X1730-333  & varies &  unknown & T \\
	X1820-303  & steady  & triple & G,B,A,R \\
	X1916-053  & varies  & superhumps & B,A,D \\
	X1957+115  & varies  & $\Delta \dot{M}$ & U \\
	\hline
	\end{tabular}	
	\smallskip \\
	\footnotesize{$^{*}$ OD01}\\
	\footnotesize{$^{[L]}$ Liu, van Paradijs \& van den Heuvel (2007) and ref. therein,}\\
	\footnotesize{A: known atoll source,}
	\footnotesize{B: X-ray burst source,}\\
	\footnotesize{D: ”dipping” low-mass X-ray binary,}\\
	\footnotesize{E: eclipsing or partially eclipsing low-mass X-ray binary,}\\
	\footnotesize{G: globular-cluster X-ray source,}\\
	\footnotesize{M: microquasar,}
	\footnotesize{P: X-ray pulsar,}\\
	\footnotesize{R: radio loud X-ray binary,}
	\footnotesize{T: transient X-ray source,}\\
	\footnotesize{U: ultra-soft X-ray spectrum,}
	\footnotesize{Z: Z-type source}\\
    \end{center}
\end{table}

\subsection{Summary of Results} 

A summary of all 25 sources is presented in Table \ref{hmxbs} \& \ref{lmxbs} for HMXBs and LMXBs respectively. Therein, concise descriptions of the behaviour of their superorbital periods are given, together with the mechanisms believed to be responsible. Relevant references regarding mechanisms are contained in the text under discussion of each source. The tables also include the source types included in the catalogue papers by Liu, van Paradijs \& van den Heuvel (2006,2007). 

Chaotic refers to unstable or highly variable periodic signals, like those displayed by Cyg X-2. Varying includes evolution of persistent periodic signals (like SMC X-1) or intermittent periodic signals (like X1636-536). Steady refers to persistent or intermittent periodic signals that display no significant variation (like Her X-1 \& LMC X-4), i.e. the variations are within conservative error estimates for the periods.

\section{Conclusions}

This DPS overview aims to provide the long awaited time-resolved analysis that OD01 suggested would be required for aperiodic sources, as a basis for future investigations. An in-depth analysis of each source and detailed discussion of the full implications of these results, goes beyond the scope of this work. Here we wish to highlight whether the results from our DPS analysis suggest steady, evolving, persistent or unstable periods, and how they relate to previously published periods.

During the discussion of each source, we first compared the observed behaviour apparent from our analysis with the OD01 predictions for stability of accretion discs against radiation-driven warping, since the OD01 process has a dependence on system parameters (determining an XRB's location on Fig.\ref{OD2001}) which allows its feasibility for superorbital varying systems as a class to be assessed. Sources for which other mechanisms can account for the $P_{sup}$, were also compared to those predictions, where applicable. Such mechanisms include tidal disc precession, precessing relativistic jets and $\dot{M}$ variations due to a third body, state changes or variations in the size of the equatorial disc surrounding a Be star donor.

\subsection{Warped/Tilted Accretion Discs} 

OD01 predicted that only a small fraction of X-ray binaries should display steady superorbital periods associated with stable, steadily precessing, radiation-driven warped accretion discs. Furthermore, they suggested that it would be even less common in LMXBs, since only those with $P_{orb}>$1 d are expected to produce warps. However, 16 of the 25 sources contained herein are LMXBs. With 114 HMXBs and 187 LMXBs contained in the catalogues of Liu, van Paradijs \& van den Heuvel (2006, 2007), it implies that superorbital periods have been detected in a slightly larger fraction of LMXBs than HMXBs.

However, recent Smooth Particle Hydrodynamics (SPH) simulations by Foulkes, Haswell \& Murray (2010) on SMC X-1, Cyg X-1, Cyg X-2, X1916-053, LMC X-3, Her X-1, SS433 and a generalized LMXB, produced warps over all orbital periods considered. Contrary to OD01, they predict that superorbital periods should be very common in LMXBs, suggesting that the analytical OD01 approach is necessarily approximate and that the SPH simulations should incorporate the complexities involved in an irradiated accretion disc more accurately as their ``disc continuously flexes in response to the changing orientation of the Roche potential''.

It is important to remember that the OD01 predictions for stability of accretion discs against radiation-driven warping, are for specific $\alpha$ and $\epsilon$ values and that the instability criteria depend largely on $\alpha$ and to a slightly lesser extent on $\epsilon$, making the predictions approximate. Different $\alpha$ and $\epsilon$ values would yield very different predictions and these values are expected to differ from system to system. Other mechanisms, such as wind-driven tilting and magnetic warping, may also produce warped/tilted accretions discs.

Furthermore, warps need to be sustained for prolonged intervals to produce steady superorbital periods (as in Her X-1 \& LMC X-4). Lodato \& Price (2010) used SPH simulations to consider the diffuse propagation of warps in viscous thin discs, where they determined the diffusion coefficient to be $\sim 1/\alpha$ (for small amplitude warps and $\alpha <$0.1), in general finding that higher viscosity leads to slower diffusion and lower viscosity to more rapid diffusion.

Of the 25 X-ray binaries considered here, 15 could be directly compared to their OD01 predictions. The results are generally consistent with their predictions.

Firstly, our results support the OD01 prediction that LMXBs are unlikely to produce stable precessing warped discs, since they are mostly located below the $r_{add}=r_{o}$ line and their accretion disks are therefore too small to become unstable against warping in the expected radiation field. Her X-1 remains the only LMXB to have a steady superorbital period that can be associated with its accretion disc, since X1820-303 is likely a triple system. Her X-1 is not a typical LMXB, since it has higher donor mass than most other LMXBs and therefore it might share some HMXB properties or be an Intermediate-Mass X-ray Binary (IMXB; see Podsiadlowski, Rappaport \& Han 2003). One of these properties appears to be its ability to produce a warped precessing disc, with its associated steady superorbital period.

Secondly, OD01 predictions implied that HMXBs are more likely to produce stable warped precessing discs, although eccentric orbits would suppresses stable radiation-driven warping. Our DPS analysis has shown that not only LMC X-4 displays a stable superorbital period, but also Cyg X-1, SS433, X0114+650 \& XTE J1716-389. Only the locations of the latter two sources on Fig.\ref{OD2001} are unknown. Cyg X-1, LMC X-4 \& SS433 are located in the OD01 instability zone and therefore expected to have warped precessing accretion discs that produce steady superorbital periods while they are persistent. However, the steady superorbital periods in SS433 and X0114+650 may have other origins, since the former experiences relativistic jet precession and the latter is a BeX.

Thirdly, we included 5 quasi-persistent sources for which superorbital periods have been published. Four of these are LMXBs (X1636-536, KS 1731-260, EXO 0748-676 \& IGR J17098-3628) and XTE J1716-389 is the only HMXB. As OD01 predicted, these quasi-persistent sources produce superorbital periods during their persistent (high) states, with the exception of X1636-536, where stronger detections occurred during the lower state. Its low state is certainly not an {\it off} state where the flux goes to zero, as seen in the other four sources. It also displayed an anti-correlation between hard and soft X-ray components (Shih et al. 2005). Furthermore, only the HMXBs produced a relatively steady superorbital period while the LMXBs all produced multiple periodic signals that were either evolving or unstable/chaotic.

Finally, OD01 predicted that Sco X-1 would be marginally unstable or stable and show variability, since it is located in the intermediate instability zone. Our results show long-term behaviour that may be interpreted in this manner.

Clarkson et al. (2003) suggested that warps may manifest themselves as modulations in the X-ray flux, not only resulting from variations in the accretion disc structure but also by modulation of $\dot{M}$. Varying absorption of X-rays from the central source by a warped inner disc or the variation in the uncovered X-ray emitting area apply to the former, while variation of the accretion rate onto the compact object or modulation of the mass flow rate through the disc (Dubus 2003) apply to the latter. It was also suggested that competing radiation-driven warping modes may cause variations in the warp itself, resulting in the steady evolution/variation of the $P_{sup}$ in SMC X-1, consistent with its OD01 prediction. Her X-1, with $P_{sup}\sim$33-37 d (Leahy \& Igna 2010) might be subject to the same effect, to a lesser extent.

\subsection{Complex $\dot{M}$ variations} 

Many of the superorbital periods already detected may be the result of the modulation of the $\dot{M}$, rather than only being associated with warped precessing accretion discs. Many may therefore also indicate the time-scale on which transitions occur between high and low flux states. The HMXBs Cen X-3, Cyg X-1 \& LMC X-3 and the LMXB X1957+115 have very similar lightcurves that clearly show these transitions on varying time-scales. However, in the case of Cyg X-1 there is also evidence for a precessing warped accretion disc (Zdziarski, Pooley \& Skinner 2011).

In fact, the majority of the superorbital periods are clearly not steady, with many of those showing unsteady evolution of multiple periodic signals while they are persistent. Such behaviour may also be associated with a variety of time-scales produced by $\dot{M}$ variations.

\subsection{The strength of time-dependent analysis} 

Our results emphasize the importance of using a time-dependent periodic analysis approach such as DPS when investigating the long-term periods that have been associated with accretion disc properties such as precession/warping/tilting and $\dot{M}$ variations due to X-ray state changes. Traditional period analysis, which use the entire datasets at one's disposal, would miss a source like KS 1731-260 since its overall periodogram shows no significant peaks. Moreover, all the variations and complex behaviour of quasi-periodic signals go completely unnoticed. For example, the lightcurves of LMXBs X1730-333 \& GRS 1747-312 show some similarity to Her X-1, but their superorbital periods are not steady and appear to be evolving or varying.

Consequently it is not unexpected that studies of superorbital behaviour in hard X-rays (14-195 keV) using data from the Swift Burst Alert Telescope and soft X-rays (1.5-12 keV) using RXTE ASM data, found no significant periodic modulations for X1916-053, Cyg X-2 \& Sco X-1, while being able to confirm the previously reported periods in X1820-303 \& X1636-536 (Farrell, Barret \& Skinner 2009). Unstable and/or intermittent periodic signals may lie below the white and/or red noise level when considering the longer datasets, while the time-dependent DPS analysis shows intermittent and sustained periodic signals that appear to vary over time. 

Since we are no longer limited by desktop technology, applying this analysis method to the other XRB sources contained in the RXTE ASM is relatively straightforward. It may reveal complex periodic behaviour that has thus far gone unnoticed.

\section{Acknowledgements}
ASM results were provided by the ASM/RXTE teams at MIT and at the RXTE Science Operations Facility and Guest Observer Facility at NASA's Goddard Space Flight Center (GSFC). 
Starlink's PERIOD package was used to produce all periodograms and window functions. All scripts to automate the analysis process were written in PYTHON.

We would like to thank Dr. Alan Levine for the use of his Fortran program to obtain the white noise estimates for the significance test.
We thank Guillaume Dubus for his comments and for pointing out the erroneous superorbital period associated with Sco X-1 in OD01. 
We also wish to thank Will Clarkson for his detailed comments and suggestions. 
Finally, we thank Brian Warner for his continued support and comments.





\label{lastpage}

\end{document}